\documentclass{article}
\usepackage{amsmath,amssymb}
\usepackage{mathtools}
\usepackage{spconf,graphicx}
\usepackage{subcaption}
\usepackage{color}
\usepackage{siunitx}
\usepackage{textcomp}
\usepackage{tabulary}
\usepackage[font={small,it}]{caption}

\title{SAPSAM - Sparsely Annotated Pathological Sign Activation Maps - A novel approach to train Convolutional Neural Networks on lung CT scans using binary labels only}

\name{M. Zusag$^{1}$, S.R. Desai$^{2,4}$, M. Di Paolo$^{5}$, T. Semple$^{2,4}$, A. Shah$^{2,5}$, E.D. Angelini$^{1,3}$}
\address{$^{1}$NIHR Imperial Biomedical Research Centre, ITMAT Data Science Group, \\Imperial College London, London, UK \\
$^{2}$NHLI, Imperial College London, London, UK \\
$^{3}$Department of Biomedical Engineering, Columbia University, New York, USA \\
$^{4}$Department of Radiology, Royal Brompton and Harefield NHS Foundation Trust, London, UK \\
$^{5}$Department of Respiratory Medicine, Royal Brompton and Harefield \\ NHS Foundation Trust, London, UK
}

\begin{document}

\maketitle
\begin{abstract}
\emph{Chronic Pulmonary Aspergillosis (CPA)} is a complex lung disease caused by infection with \emph{Aspergillus}. Computed tomography (CT) images are frequently requested in patients with suspected and established disease, but the radiological signs on CT are difficult to quantify making accurate follow-up challenging. We propose a novel method to train Convolutional Neural Networks using only regional labels on the presence of pathological signs, to not only detect CPA, but also spatially localize pathological signs. We use average intensity projections within different ranges of Hounsfield-unit (HU) values, transforming input 3D CT scans into 2D RGB-like images.  CNN architectures are trained for hierarchical tasks, leading to precise activation maps of pathological patterns. Results on a cohort of 352 subjects demonstrate high classification accuracy, localization precision and predictive power of 2 year survival. Such tool opens the way to CPA patient stratification and quantitative follow-up of CPA pathological signs, for patients under drug therapy.

\end{abstract}

\section{Introduction}
\label{sec:intro}
Chronic Pulmonary Aspergillosis (CPA) is caused by infection with Aspergillus species, which are ubiquitous in nature. CPA most often occurs in patients with pre-existing lung disease. Predisposing host systemic  for the development of CPA include diabetes, alcoholism, a low body mass index and cancer. Typical patients with CPA have non-specific insidious symptoms for over 3 months including weight loss, cough, occasional haemoptysis and low-grade fever. On imaging, disease is most frequent in the upper lobes and involves pathological signs such as consolidation, cavity formation, volume loss and striking pleural thickening \cite{RefWorks:doc:cpa}. Ancillary signs on CT include nodules, large airway mucus plugging and a "tree-in-bud" pattern, the latter reflecting inflammation in the peripheral airways. 
Multiple recent approaches exploit convolutional neural networks (CNNs) and transfer learning to learn pathological signs on lung CT scans \cite{RefWorks:doc:nam}. Given the challenge of manually annotating voxel-wise (or even slice-wise) 3D volumes of images, it is of value to train CNN labelers on whole volumes with binary labels only (diseased/non-diseased), while being able to feedback to the user the spatial locations that were used to make the labeling decision. As CT volumes are too large to fit on common GPUs, most approaches either work on downsampled 2D slices or 3D/2D patches, like \cite{RefWorks:doc:3dcnn}. We chose to test an alternative radical image simplification strategy, projecting 3D volumes into 2D images. The rationale for using such strategy is five-folds: (1) Require whole-scan labels only; (2) Use an image simplification scheme similar to the maximum intensity projections popular for some visualization tasks; (3) Test CNN capacities to infer fine diagnostic tasks on projected image data, mimicking radiologists capabilities when reviewing 2D lung X-rays; (4) Exploit the plethora of available CNN architectures pre-trained for 2D image classification and labeling, while using deep network architectures; (5) Reduce the ratio of pathological signs to other lung tissues, and improve the balance in the training data.

\subsection{Data}
A retrospective cohort of 352 high-resolution CT studies from 172 patients was collected under an ethically-approved study protocol. The CT scans were acquired at full inspiration, using 120 kVp, axial pixel size between 0.44mm and 0.9mm and slice thickness between 0.5mm and 1.0mm. Working with the lowest pixel resolution of 0.9mm and slice thickness of 1mm, our CT scans consist, on average, of $512 \times 512 \times 432$ pixels. The cohort includes 76 control studies, without any underlying lung disease and 276 studies with signs of CPA, taken from N=96 patients who had between 1 and 8 longitudinal CT scans acquired over a period of 12 years. Pre-existing (non-exclusive) lung diseases in this cohort are: 112 cases with consolidation, 86 cases with bronchiectasis, 66 cases with emphysema, 39 cases with sarcoidosis, 38 cases with signs of ground glass opacities, 26 cases with cystic fibrosis and 31 cases with a disease not in this list. The three pathological signs of CPA being studied in this work and illustrated in Fig. \ref{fig:axial-projs} are: cavities, fungus balls and pleura thickening.

\subsection{CT scan manual annotation}
All CPA CT scans were visually inspected by an expert radiologist and sparsely annotated by dividing each CT scan into 6 sub-regions (left and right upper/middle/lower regions) and indicating, for each sub-region, if any of the three pathological signs and if a pre-existing lung condition could be identified. Each CT scan is therefore equipped with 6 one-hot encoded vectors of length 4 containing, the manual annotation on the presence or absence of any pre-existing lung condition and each of the 3 pathological signs being studied. Occurrences, in our cohort, of CPA pathological signs per sub-regions are detailed in Table \ref{table:occurence-pathological-signs}.
We see that CPA pathological signs occur most often in the upper lung, which creates some imbalance between sub-regions. 
% and that pathological signs tend to the manual labels from the CPA CT scans are very imbalanced.

\begin{table}[htb!]
\caption{Occurrences of pathological signs in our cohort of subjects with CPA (N=276 studies and 1,656 sub-regions).}
\centering
%\begin{tabular}{|l|c|c|c|}
%\begin{tabular}{|p{2cm}|p{1.5cm}|p{1.5cm}|p{1.5cm}|}
\begin{tabular}{|p{2cm}|c|c|c|}
\hline
 \textbf{\begin{tabular}[c]{@{}l@{}} \# scans \end{tabular} \newline \begin{tabular}[c]{@{}l@{}} (left \textbar right) \end{tabular}} & \textbf{\begin{tabular}[c]{@{}l@{}} Cavities\end{tabular}} & \textbf{\begin{tabular}[c]{@{}c@{}} Pleura \\ thickening\end{tabular}} & \textbf{\begin{tabular}[c]{@{}c@{}}  Fungus \\ balls\end{tabular}}   \\ \hline
\textbf{\begin{tabular}[c]{@{}l@{}} Upper lung\end{tabular}} & 172 \textbar \ 121 & 169 \textbar \ 143 & 119 \textbar \ 143 \\ \hline
\textbf{\begin{tabular}[c]{@{}l@{}}Middle lung\end{tabular}} & 33 \textbar \ 15 & 12 \textbar \ 8 & 14 \textbar \ 10  \\ \hline
\textbf{\begin{tabular}[c]{@{}l@{}}Lower  lung\end{tabular}} & 5 \textbar \ 15 & 8 \textbar \ 3 & 3 \textbar \ 9   \\ \hline
 \end{tabular}
\label{table:occurence-pathological-signs}
\end{table}
\vspace{-1.5em}

%=======================================
\section{Method}
\label{sec:methods}
\begin{figure*}[ht!]
    \centering
    \includegraphics[width=0.98\textwidth]{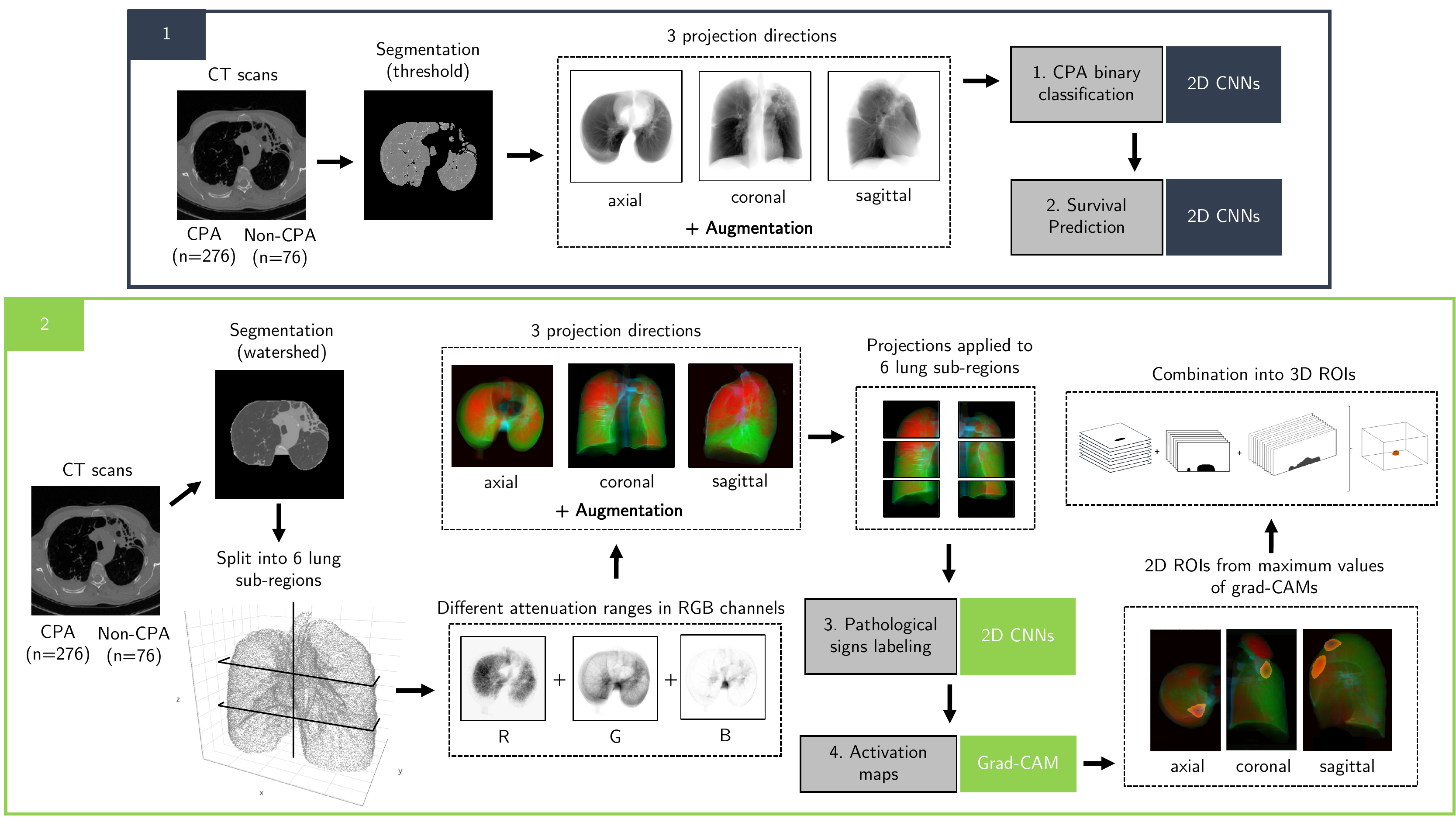}
    \caption{Pipeline of SAPSAM framework: (1) binary disease classification and survival prediction on whole-scan projections along each axis; (2) Sub-regions pathological signs labeling using different attenuation ranges in RGB-like channels.}
    \label{fig:pipeline}
\end{figure*}

The pipeline of the proposed framework is displayed in Fig. 
\ref{fig:pipeline}, and includes two distinct classification tasks: (1) whole-scan binary CPA disease classification and survival prediction; (2) sub-regions pathological signs labeling. 

\subsection{Lung segmentation}
\label{ssec:segmentation}
\begin{figure}[hb!]
  \begin{subfigure}[t]{0.24\linewidth}
    \includegraphics[width=\textwidth]{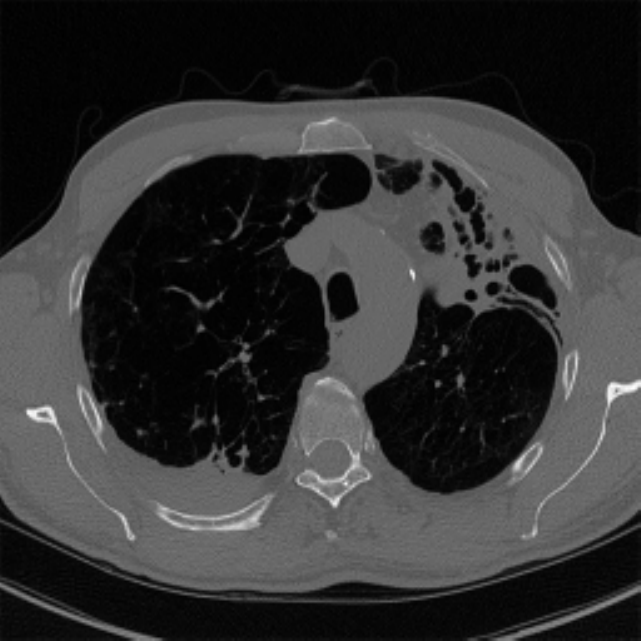}
    \caption{Axial slice}
    \label{fig:original-watershed}
  \end{subfigure}\hfill
  \begin{subfigure}[t]{0.24\linewidth}
    \includegraphics[width=\textwidth]{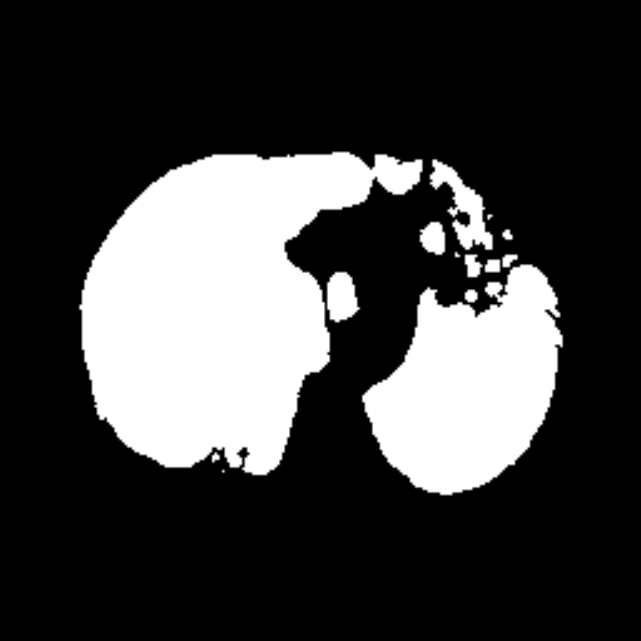}
    \caption{Watershed}
    \label{fig:watershed}
  \end{subfigure}\hfill
  \begin{subfigure}[t]{0.24\linewidth}
    \includegraphics[width=\textwidth]{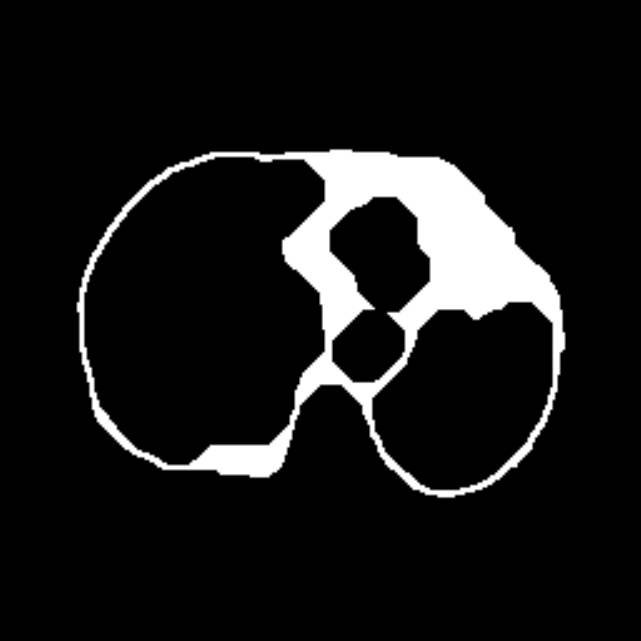}
    \caption{Outline}
    \label{fig:outline}
  \end{subfigure}\hfill
  \begin{subfigure}[t]{0.24\linewidth}
    \includegraphics[width=\textwidth]{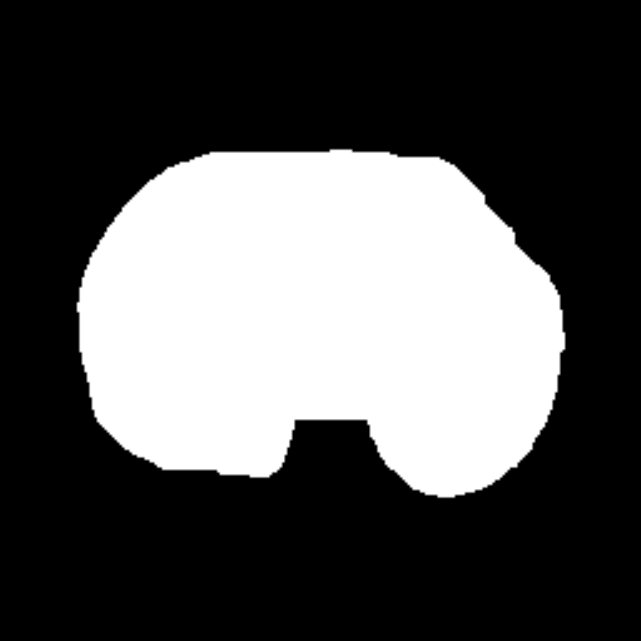}
    \caption{Closing}
    \label{fig:closing}
  \end{subfigure}\hfill
  \vspace{-.5em}
  \caption{Illustrations of the segmentation of the lung area}
  \label{fig:segmentation-procedure}
\end{figure}

For the binary disease classification/survival prediction  tasks,
we used a conservative approach, thresholding the voxels at -570 HU and using morphological closing and opening operations to create smooth lung masks that contain lung parenchyma and air. 
To further include pathological signs such as fungus balls and pleura-thickening, we refined the approach from \cite{RefWorks:doc:watershed-paper} and \cite{RefWorks:doc:watershed}. First, we define a set of markers. To do so, pixels are thresholded at -570 HU and connected components are extracted, while removing the smallest ones. These connected components provide internal markers. Then, we define intermediate and external markers, via morphological dilation of internal markers with structuring elements of radius 10 and 35 pixels, respectively. These markers define the spatial extent of the region where lung tissues can be added to the threshold-based segmentation. To find the lung structures to add, a watershed segmentation (Fig. \ref{fig:watershed}) is used, with seeds generated with a Sobel-filter edge map. It returns  large homogeneous connected  regions but still excludes some diseased structure with soft-tissue like attenuation values. A top-hat transform (Fig. \ref{fig:outline}) is used, where the occluded region is initially morphologically closed and then the difference between the original and the closed structure is added to the watershed mask. The final segmentation mask (Fig. \ref{fig:closing}) is generated after closing the remaining holes. 

\subsection{2D average intensity projections}
\label{ssec:projections}
 We use average intensity projections within the segmented lungs to  capture 3D pathological signs in 2D images. Since the pathological signs of interest exhibit  different attenuation levels (air cavities, normal lung parenchyma, soft tissue), we sub-divided attenuation values into three ranges, to create three RGB-like channels, using: [-1,400HU, -900HU] (air-like range), [-900HU,-160HU] (lung range), [-160HU, 240HU] (soft-tissue range). 
Similar partitioning of attenuation values was used in \cite{RefWorks:doc:holistic-classification} to classify 6 subtypes of lung disease.

\subsection{CT projection and augmentation}
The scans were all rescaled to $512 \times 512 \times 432$ pixels and downsampled by 2 in each direction. To handle the imbalance issues, we augmented our cohort by rotating each scan in 3D by $\pm 5$ degrees around each axis, i.e. 27 rotations in total. 
%Scans were segmented to extract lung regions using the thresholding or watershed approach as indicated in Fig. \ref{fig:pipeline}.
For the binary disease classification/survival prediction tasks, each scan was projected along the three orthogonal axis leading to three gray-scale projection images. Intensities in projection images were rescaled to  [0,1]. For the sub-region labeling tasks, each scan was subdivided into 6 sub-regions of size $256 \times 128 \times 72$ voxels. Each sub-region was projected along the three orthogonal axes and within the 3 ranges of HU values, leading to three color projection images. Intensities within each channel in the projection images were rescaled to  [0,1]. Finally, the projection images were randomly rotated by $\pm 20$ degrees and randomly scaled between $80\%$ and $120\%$ of their original size, which resulted in several 100,000 of train/test/validation 2D images. To get more reliable measurements of the networks' generalization performance, train and test subsets were split by patients, such that a given patient cannot occur in both sets. To avoid any bias of the network towards one class, train and test subsets were balanced after the split.

\subsection{CNN architectures}
\label{ssec:DL-bin-class}
We trained the VGG19 \cite{RefWorks:doc:vgg19} and InceptionV3 \cite{RefWorks:doc:inception} architectures for binary disease classification on all 3 projection directions, resulting in 12 CNNs. With the VGG19 architecture performing best for this task, we only experimented with this architecture for pathological signs detection in all 3 projection directions, leading to 9 CNNs ( for 3  pathological signs), and 21 CNNs trained in total. For both architectures, we loaded the pre-trained weights from the ImageNet competition and discarded the final fully-connected layer, which was optimized to label 1,000 classes from ImageNet. We replaced it with 2 fully connected layers of size $1,024 \times 1 \times 1$, where the weights were randomly initialized by sampling from a Gaussian distribution with biases set to 0. Both layers use a rectified linear activation function and a dropout between them of 0.5. The output layer was set to a fully connected layer with 2 outputs using the softmax function for the final output activation. 

For the survival prediction task, the VGG19 CNN architecture optimized for the binary disease classification task was used as a base model. The last two fully-connected layers were reset to a random state, and re-trained.

\subsection{PSAMs - 3D Pathological Sign Activation Maps}
\label{ssec:psams}
We use the concept of gradient-weighted class activation maps (grad-CAM) by Selvaraju et al. \cite{RefWorks:doc:grad-cam}, to visualize regions with high activations of the softmax outputs for the pathological signs classes. They are based on the idea that convolutional layers develop localization capabilities of objects they were trained on, as initially suggested in  \cite{RefWorks:doc:object-detectors}. Mathematically, for a specific target class $c$, we note $y^c$ as the final labeling score and  $A^{k} \mid k \in \{1, \dots, n \}$ as the feature maps of the final convolutional unit. The grad-CAM $L^c$ for class $c$ is defined:

\begin{align}
L^{c} = ReLU\left (\sum_{k=1}^{n} \alpha_{k}^{c} A^{k}\right ) \mid
\alpha_{k}^{c} = \frac{1}{Z} \sum_{i=1}^{u}\sum_{j=1}^{v} \frac{\delta y^{c}}{\delta A_{i,j}^{k}}
\end{align}
\noindent where $A^{k} \in \mathbb{R}^{u\times v}$ and $Z \in \mathbb{R}$.
The output of the linear combination of the weight parameters $\alpha_{k}^{c}$ and the feature maps $A^{k}$ is rectified so that only the feature maps that have a positive influence on the target class $c$ are visualized. 
We evaluate if the grad-CAMs  can yield accurate spatial localization information from CNNs trained for pathological signs labeling on intensity projections of lung sub-regions.

%=======================================================
\section{Results}
\label{sec:results}

\subsection{Binary disease classification}
\label{ssec:Res-bin-class}
We achieved the best generalization performance on N=822 \textbf{axial} projections from the test set with \textbf{precision=97.4\%}, \textbf{recall=94.8\%} and \textbf{$\text{F}_1$-score=96\%}, measured on the categorical cross-entropy loss over balanced subsets. The InceptionV3 had slightly inferior test results in our experiments (\textbf{precision=95.8\%}, \textbf{recall=93.6\%} and \textbf{$\text{F}_1$-score=94.7\%}. The best VGG19 model architecture uses the following hyperparameters: \textbf{optimizer} = Stochastic Gradient Descent, \textbf{momentum} = 0.9, \textbf{learning rate} = $10^{-4}$, \textbf{frozen layers} = the first two blocks, \textbf{batch size} = 32, \textbf{epochs} = 60, \textbf{callbacks} = checkpoint evaluated on the validation data, \textbf{rotation factor} = 20 degrees , \textbf{zoom factor} = 0.1.

\subsection{Sub-regions disease sign labeling }
\label{ssec:Res-lat-class}
The best optimizer, according to the categorical cross-entropy loss function from the VGG19 network, was obtained using: \textbf{optimizer} = Stochastic Gradient Descent, \textbf{momentum} = 0.9, \textbf{learning rate} = $10^{-4}$, \textbf{frozen layers} =  first two blocks, \textbf{batch size} = 32, 
\textbf{epochs} = 240-300,
\textbf{callbacks} = checkpoint evaluated on the validation data, 
\textbf{rotation factor} = 20 degrees, \textbf{zoom factor} = 0.1. 
After balancing, we used ~13,000 training and 3,300 testing projection images. Labeling test-scores are reported in Table \ref{table:results-all-networks}.

\begin{table}[h!]
\caption{Test scores (\%) of region-based labeling of pathological signs on  Axial \textbar Coronal \textbar Sagittal projections.}
\centering
\begin{tabular}{|l|c|c|}
\hline
\textbf{\begin{tabular}[c]{@{}c@{}} Labeling Task \end{tabular}} & \textbf{\begin{tabular}[c]{@{}c@{}} Precision \end{tabular}} & \multicolumn{1}{c|}{\textbf{\begin{tabular}[c]{@{}c@{}}Recall\end{tabular}}} \\ 
\hline
\textbf{\begin{tabular}[c]{@{}l@{}} Cavities \end{tabular}} & \textbf{89.2} \textbar 85.3 \textbar 86.3 & \textbf{90.0} \textbar 88.5 \textbar 87.5 \\ 
\hline
\textbf{\begin{tabular}[c]{@{}l@{}}Fungal ball \end{tabular}} & \textbf{86.3} \textbar 84.2 \textbar 82.0 & \textbf{81.9} \textbar 80.4 \textbar 76.3\\ 
\hline
\textbf{\begin{tabular}[c]{@{}l@{}}Pleura thickening \end{tabular}} & 90.6 \textbar \textbf{91.5} \textbar 78.1 & \textbf{95.2} \textbar 92.7 \textbar 82.5 \\ 
\hline
\end{tabular}
\label{table:results-all-networks}
\end{table}

\begin{figure}[hb!]
\centering
    \includegraphics[width=0.47\textwidth]{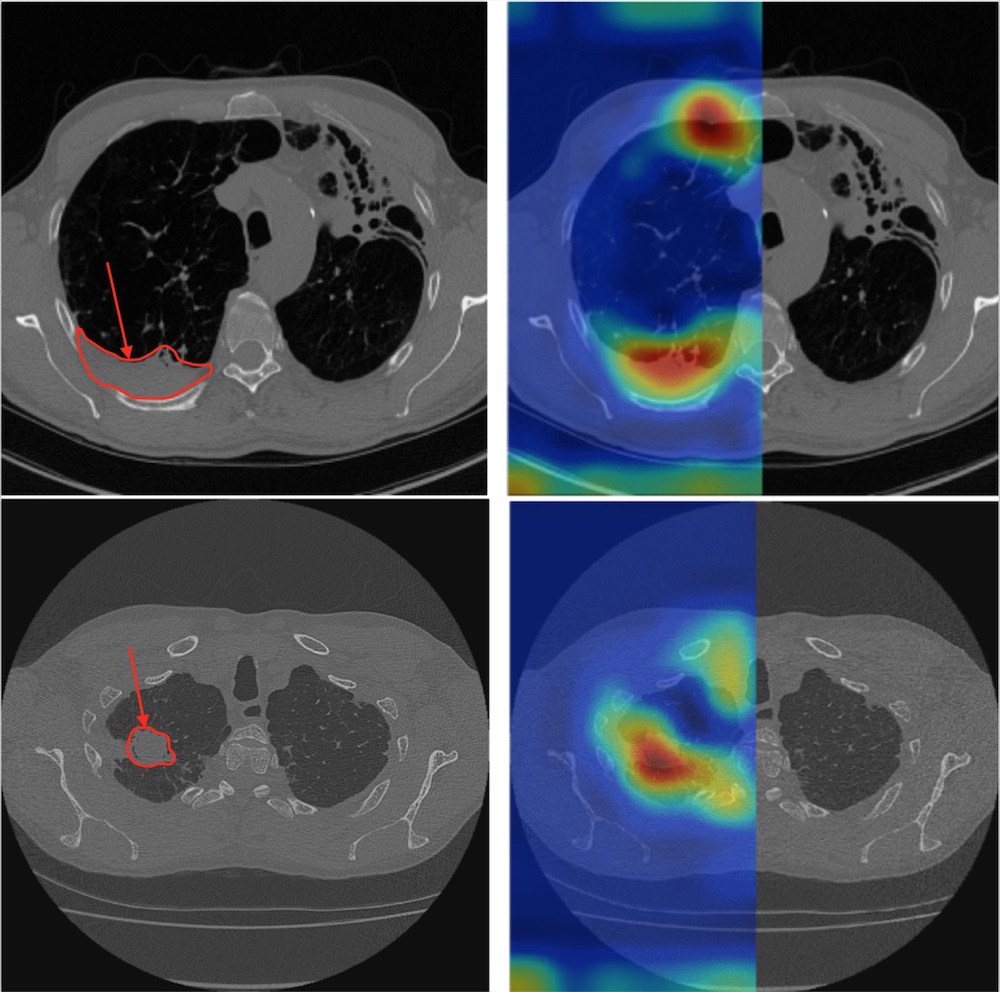}
    \caption{Localization of pathological signs using axial projections on CPA subjects with pleura thickening (top) and a fungus ball (bottom). The original axial slice (left) and the overlaid grad-CAMs (right) are shown.}
  \label{fig:axial-projs}
\end{figure}

% The networks trained on the \textbf{axial} views achieve the highest test accuracy for the three pathological signs. 
\noindent Test scores are best or close to best using \textbf{axial} views, for the three pathological signs. 
We illustrate in Fig. \ref{fig:axial-projs} two axial Grad-CAM activation maps, where 
We illustrate in Fig. \ref{fig:axial-projs} two axial Grad-CAM activation maps. In both figures, an original CT slice from a CPA patient is shown along with the gradient weighted activation map from the corresponding sub-region. We can observe that locations of maximum activation match very well with actual locations of pleura thickening and fungus balls.

\subsection{Survival prediction}
\label{ssec:DL-survival}
We tested the extension of the binary disease classifier to predict survival of CPA patients within 2 years from scanning time. 
To do so, we gathered the following scans from a subset of (N=143) CPA subjects: 
For (N=43) dead CPA subjects, we gathered all the CT scans that had been acquired within the preceding two years.
We assigned  a positive label to all these CT scans. 
For (N=100) CPA subjects, we gathered all scans that were acquired more than 2 years prior to the date of annotation, and assigned those cases a negative label. 
This lead to N=765 projections, divided into (N=510) training cases and (N=255) testing cases, without overlap of subjects between the two sets. 
No validation set was used, due to the small sample size available. The model was trained for 20 epochs as more epochs led to over-fitting. We achieved the following test-scores: \textbf{precision=81.7\%}, \textbf{recall=83.6\%} and \textbf{$\text{F}_1$-score=82.6\%}.
\section{Conclusion}
\label{sec:conclusion}
We propose an original deep-learning framework for lung CT scans,  trained using only sparse (approximate) annotations of disease state and presence of pathological signs, to enable disease classification, pathological signs localization and 2-year survival prediction.  
One original component is the use of average intensity projections of segmented lung tissue to capture 3D contextual information in 2D, while limiting computational complexity. Future work will focus on 3D positioning of the detected pathological structures. 

% \vfill\pagebreak
\newpage
\nocite{RefWorks:doc:lung-segm}
\nocite{RefWorks:doc:treatment-cpa}
\nocite{RefWorks:doc:lung-nodules}
\nocite{RefWorks:doc:visualizing-cnns}
\nocite{RefWorks:doc:imagenet}
{
\bibliographystyle{IEEEbib}
\bibliography{strings,refs}
}
\end{document}